\documentclass[aps,reprint,prb,showpacs]{revtex4-1}
\usepackage{natbib}
\usepackage[english]{babel}
\usepackage{amsfonts,amsmath,amssymb,bm}
\usepackage[fleqn,tbtags]{mathtools}
\usepackage{graphicx}
\usepackage[usenames,dvipsnames]{xcolor}

\usepackage[draft,markup=default,authormarkuptext=name]{changes}
\definechangesauthor[name={Win},color={Red}]{w}
\definechangesauthor[name={Andreas},color={orange}]{ak}
\definechangesauthor[name={Ara},color={blue}]{as}
\definechangesauthor[name={Ilya},color={green}]{i}
\definechangesauthor[name={Noah},color={cyan}]{n}

\usepackage{hyperref}

\usepackage{comment}

\allowdisplaybreaks[1]

\newcommand{\be}{\begin{equation}}
\newcommand{\ee}{\end{equation}}
\newcommand{\bes}{\begin{equation}\begin{split}}
\newcommand{\ees}{\end{split}\end{equation}}
\newcommand{\bea}{\begin{eqnarray}}
\newcommand{\eea}{\end{eqnarray}}

\def\beq{\begin{equation}}
\def\eeq{\end{equation}}
\def\bea{\begin{eqnarray}}
\def\eea{\end{eqnarray}}

\begin{document}

\title{Critical behavior at the integer quantum Hall transition in a network model on the Kagome lattice}

\author{I. A. Gruzberg}
\affiliation{Ohio state university, Department of Physics, 191 West Woodruff Ave, Columbus OH, 43210}

\author{N. Charles}
\affiliation{Ohio state university, Department of Physics, 191 West Woodruff Ave, Columbus OH, 43210}

\author{A. Kl\"umper}
\affiliation{Wuppertal University, Gaußstraße 20, 42119 Wuppertal, Germany}

\author{W. Nuding}
\affiliation{Wuppertal University, Gaußstraße 20, 42119 Wuppertal, Germany}

\author{A. Sedrakyan}
\affiliation{Alikhanian National Laboratory,
	Yerevan Physics Institute, Br. Alikhanian 2, Yerevan 36, Armenia}

\begin{abstract}

We study a network model on the Kagome lattice (NMKL). This model generalizes the Chalker-Coddington (CC) network model for the integer quantum Hall transition. Unlike random network models we studied earlier, the geometry of the Kagome lattice is regular. Therefore, we expect that the critical behavior of the NMKL should be the same as that of the CC model. We numerically compute the localization length index $\nu$ in the NKML. Our result $\nu= 2.658 \pm 0.046$ is close to CC model values obtained in a number of recent papers. We also map the NMKL to the Dirac fermions in random potentials and in a fixed periodic curvature background. The background turns out irrelevant at long scales. Our numerical and analytical results confirm our expectation of the universality of critical behavior on regular network models.
	
\end{abstract}


\date{March 13, 2020}

\maketitle


{\it Introduction.} The integer quantum Hall (IQH) tran\-si\-ti\-on \cite{Huckestein-Scaling-1995} is a quantum phase transition accompanied by universal critical phenomena. A central characteristic of the transition is the exponent $\nu$ describing the divergence of the localization length of single-particle wave functions with energies $E$ close to critical energies $E_c$:
\begin{align}
\label{exponent-nu-def}
\xi \sim |E - E_c|^{-\nu}.
\end{align}
Multiple experiments \cite{Wei-Experiments-1988, Koch-Experiments-1991, Koch-Size-dependent-1991, Koch-Experimental-1992, Engel-Microwave-1993, Wei-Current-1994, Li-Scaling-2005, Li-Scaling-2009, Giesbers-Scaling-2009} demonstrated scaling near the integer QH transition in various systems. All experiment seem to be consistent with the value $\nu_{\text{exp}} \approx 2.38$ (with an important caveat, see a discussion in Ref. \onlinecite{Pruisken-Comment-2009}).

The QH plateaus separated by the transition are successfully described by models of non-interacting electrons in the presence of disorder. In this approximation the transition is an Anderson transition \cite{Evers-Anderson-2008}. Even with this simplification the problem of the IQH transition is notoriously difficult. A notable proposal for a conformal field theory of the transition  \cite{Zirnbauer-The-integer-2019} predicts logarithmic (as opposed to power-law) scaling effectively meaning $\nu = \infty$.

There are many numerical simulations of non-interacting models of the IQH transition. One of the better studied models is the Chalker-Coddington (CC) network model on a square lattice \cite{Chalker-Percolation-1988,Kramer-Random-2005}. Recent accurate simulations of the CC model \cite{Slevin-Critical-2009, Obuse-Conformal-2010, Amado-Numerical-2011, Obuse-Finite-2012, Slevin-Finite-2012, Nuding-Localization-2015} give the value $\nu$ in the range 2.56--2.62, which is definitely different from the experimental value. Similar values  have been obtained in numerical simulations of other non-interacting models of the IQH transition \cite{Dahlhaus-Quantum-2011, Fulga-Topological-2011, Zhu-Localization-2018, Puschmann-Integer-2018}.

The likely source for the discrepancy between the experimental and numerical values of $\nu$ are the electron-electron interactions \cite{Lee-Effects-1996, Wang-Short-range-2000, Burmistrov-Wave-2011}. Recently we have proposed another possible reason for the discrepancy, and studied a version of the network model on random graphs \cite{Gruzberg-Geometrically-2017, Kluemper-Nuding-Sedrakyan-2019}. Our results suggest that the additional geometric randomness is relevant: it changes the localization length exponent to $\nu \approx 2.37$ and places the random network model in a different universality class than the regular CC model.

Random graphs that we have studied are dual to random quadrangulations. A polygon with $n$ sides in a random graph is dual to a vertex where $n$ quadrangles meet. If all quadrangles are viewed as squares, the deficit angle at the vertex is $R_n = (4-n)\pi/2$, and this can be interpreted as discrete curvature of a conical singularity at the vertex. Random networks contain randomly placed curvatures, and averaging over geometric randomness can be interpreted as integration over configurations of quenched random gravitational field. One can study any 2d model with interacting matter fields defined by an evolution operator ($R$-matrix) on random quadrangulated surfaces \cite{Ambjorn-Matrix-2015}. This is similar to the studies of the critical 2d minimal models coupled to quantum gravity on triangulated random surfaces \cite{Kazakov-Exactly-1988, Kazakov-Recent-1988, Kazakov-Percolation-1989, Duplantier-Geometrical-1990}.

Our previous results \cite{Gruzberg-Geometrically-2017, Kluemper-Nuding-Sedrakyan-2019} raise the issue of universality of critical behavior of network models. We expect that if the network is not random but contains periodically placed fixed curvatures then the exponent $\nu$ should be the same as in the CC model. One such network can be defined on the Kagome lattice which contains triangles with $R_3=\pi/2$ and hexagons with $R_6=-\pi$ in a periodic arrangement. In this paper we study the network model on the Kagome lattice (NMKL) analytically and numerically, determine its critical behavior, and confirm our expectation of the universality of the exponent $\nu$.

\begin{figure}
	\centering
	\includegraphics{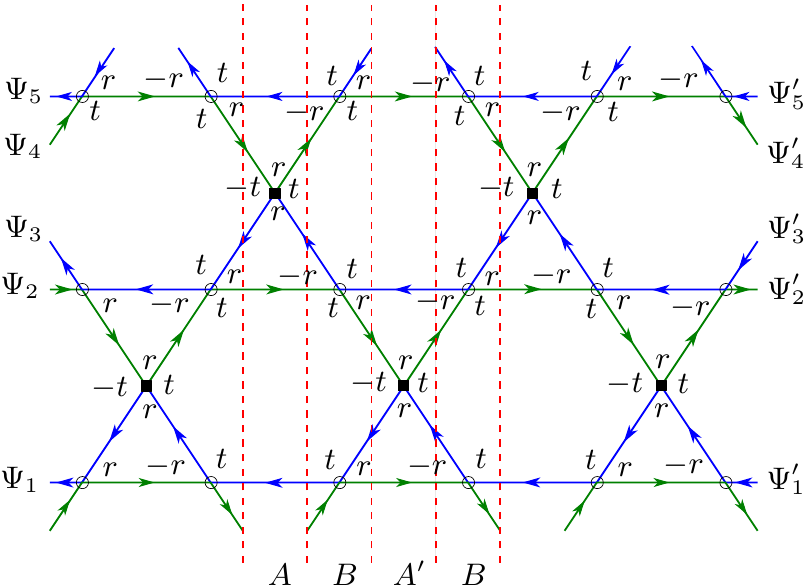}
	\caption[Kagome lattice]{The network model on the Kagome lattice. Left running channels are shown in blue, right running channels in green. The column transfer matrices $A$, $B$ and $A'$ are framed by red dashed lines. There are two different types of vertices, $a$ and $b$, shown as squares ($a$) and circles ($b$).}
	\label{fig:pic_kagome_lat}
\end{figure}

{\it The model}. The NMKL is shown in Fig.~\ref{fig:pic_kagome_lat}. A state of the network $|\psi\rangle = \sum_{l} \psi_l |l\rangle$ is a vector in ${\cal H} = \mathbb{C}^{N_L}$, where $N_L$ is the number of links, and $|l\rangle$ are basis vectors associated with each link $l$.

States of the network evolve in discrete time, each time step described by a unitary matrix $\cal U$ acting on $\cal H$ whose matrix elements ${\cal U}_{ll'}$ are non-zero only if $l'$ and $l$ are incoming and outgoing links at the same node. In this case ${\cal U}_{ll'} = e^{i\phi_{l}} {\cal S}_{ll'}$, where $\phi_{l}$ are random phases uniformly distributed on $[0,2\pi)$, and the scattering matrix ${\cal S}$ depends on the type of node (square or circle in Fig.~\ref{fig:pic_kagome_lat}):
\begin{align}
{\cal S}_a
&= \begin{pmatrix} -t & r \\
r & t \end{pmatrix},
&
{\cal S}_b
&= \begin{pmatrix} -r & t \\
t & r \end{pmatrix}
\end{align}
This choice assigns probabilities $t^2$ for all right turns and $r^2 = 1-t^2$ for all left turns.

The clean model where all $\phi_l = 0$ is periodic and is easily solved in the momentum space, see the appendix~\ref{clean}. We find that at the critical point of the clean model $t_0 = \sqrt{3}/2$ ($r_0 = 1/2$), the spectrum of the quasi-energy $\varepsilon = -i \ln {\cal U}$ contains a gapless Dirac cone, so the long-distance description is the 2D Dirac fermion. In analogy with the analysis of Refs.~\cite{Ho-Models-1996, Gruzberg-Geometrically-2017}, the addition of weak randomness in the phases $\phi_l$ leads to the field theory of a Dirac fermion with random mass, coupled to random scalar and vector potentials, and a fixed periodic curvature background. The periodic nature of the curvature background on the lattice scale makes it irrelevant in the long distance limit, and leads to the same model of Dirac fermions as in Ref.~\cite{Ho-Models-1996} for the CC model. This enforces our expectation that the critical behavior of the NMKL is the same as that of the CC model.


{\it Numerical procedure.} To compute critical exponents of the NMKL we use the transfer-matrix method \cite{mackinnon1981scaling, mackinnon1983scaling}. For finite networks of length $L$, with $M$ channels in each direction, and periodic boundary conditions in the transverse direction, we compute the product $T_L=\prod_{j=1}^L A U_{1j}B U_{2j}A' U_{3j}B U_{4j}$ of transfer matrices for $L$ layers. Each layer is split into four sub-layers, as indicated in Fig.~\ref{fig:pic_kagome_lat}. The $2M \times 2M$ transfer matrices for the sub-layers, $A$, $A'$, and $B$, contain $2\times 2$ matrices $a$ and $b$ for the scattering nodes, and $2\times 2$ identity matrices $1_2$:
\begin{align}
&\left.\begin{aligned}
        A &= \text{diag}(1_2, a, \ldots, 1_2, a), \\
        A' &= \text{diag}(a, 1_2, \ldots, a, 1_2),
       \end{aligned}
 \right.
 \quad
B = \begin{pmatrix}
       1/t & 0 & \cdots & r/t \\
       0 & b & \cdots  & 0 \\
       \vdots & \vdots & \ddots & \vdots \\
       r/t & 0 & \cdots & 1/t
     \end{pmatrix},
\nonumber\\
&a = \begin{pmatrix} 1/r & t/r \\ t/r & 1/r
\end{pmatrix},
\quad
b = \begin{pmatrix} 1/t & r/t \\ r/t & 1/t
\end{pmatrix}.
\end{align}
In addition, the random phases $\phi_l$ are combined into diagonal matrices $U_{ll'}=\exp{(i\phi_l)}\,\delta_{ll'}$.

The transmission and reflection amplitudes $t$ and $r$ at each node are shown in Fig.~\ref{fig:pic_kagome_lat}. We parametrize them as
\begin{align} \label{rt}
  r &= \big(1 + 3 e^{2x} \big)^{-1/2}, & t &= \big(1 + \tfrac{1}{3}
      e^{-2x} \big)^{-1/2}.
\end{align}
Here $x=0$ corresponds to the critical point of the clean model without randomness.
This parameterization resembles that traditionally used for the CC model. However, in the latter case there is a symmetry with respect to rotations by 90 degrees that results in the invariance of the spectrum of $T_L$ upon exchange $x \leftrightarrow -x$ or, equivalently, $t \leftrightarrow r$ even in the presence of random phases. In the NMKL there is no such symmetry, and the critical point in the random model is not expected to be at $x=0$.

$T_L$ is a product of random matrices. According to Oceledec's
theorem~\cite{oseledec1968multiplicative} the Lyapunov exponents (LEs) defined as the eigenvalues of $\log[T_L^{\vphantom \dagger} T_L^\dagger]/2L$, tend to non-random values as $L \to \infty$. The smallest positive LE $\gamma$ is inversely proportional to the localization length in the quasi-1D system with width $M$. The product $\Gamma = \gamma M$ (the ``dimensionless'' LE) becomes a universal quantity in the limit $M \to \infty$ at the critical point of the network model. In practice, a finite-size-scaling analysis relates $\Gamma$ to critical exponents of the NMKL. In addition, Tutubalin's theorem \cite{tutubalin1965limit} states that for finite systems with $L \gg 1$, the LEs have Gaussian distributions with variance $\sim (M/L)^{1/2}$. If we consider an ensemble of $N$ random networks, the variance decreases to $\sim (M/L N)^{1/2} $. Therefore, our strategy is to consider large numbers of long systems to create ensembles of $\gamma$ that have distributions close to Gaussian.

In this work we used networks of length $L=5\times 10^6$ and created ensembles of LEs $\gamma$ labeled by $a = 1, \ldots, N_\text{ens}$, where $N_\text{ens} = 200$ is the number of pairs $(x,M)_a$ that we used. The widths $M$ take 10 values $M=20, 40, ..., 200$, and the 20 values of $x$ in the range $[0.24, 0.3]$ were chosen adaptively to get more data points in the vicinity of the (a priori unknown) critical point $x_c$ (which we estimate to be $x_c=0.268$). The numbers $N_a$ of LEs in each ensemble are given in table \ref{Table:N-ensembles} in the appendix~\ref{ensembleStats}, most of them are $N_a = 624$. The total number of LEs in all ensembles is $N_\text{LE} = 130896$.

Computing large products $T_L$ directly is not possible, as many entries of the products grow exponentially with $L$. This problem is often overcome using the QR decomposition \cite{mackinnon1981scaling, mackinnon1983scaling, vonBremen1997}, where matrices $T$ in the product are decomposed as $T = Q R$ with unitary matrix $Q$ and upper right triangular matrix $R$. An alternative is to use the LU decomposition $T=PL\,U$ using a lower triangular matrix $L$ with unit diagonal, a permutation matrix $P$ and an upper triangular matrix $U$, see Ref.~\cite{numerical_recipes} for details. Simulations with the LU decomposition are about two times faster than those with the QR decomposition.

We have generated pairs of large ensembles of LEs $\gamma$ for multiple pairs $(x,M)$ using both the QR and the LU decompositions and created a histogram for each ensemble. The histograms are very well described by normal distributions as confirmed by Gaussian fits. The centers of the Gaussian peaks in a pair corresponding to the ensembles generated by the QR and the LU decompositions differ by orders of magnitude less than the peaks widths. The widths of peaks in each such pair agrees with the same precision as the centers of the peaks do.

\begin{figure}[t]
	\centering
	\includegraphics{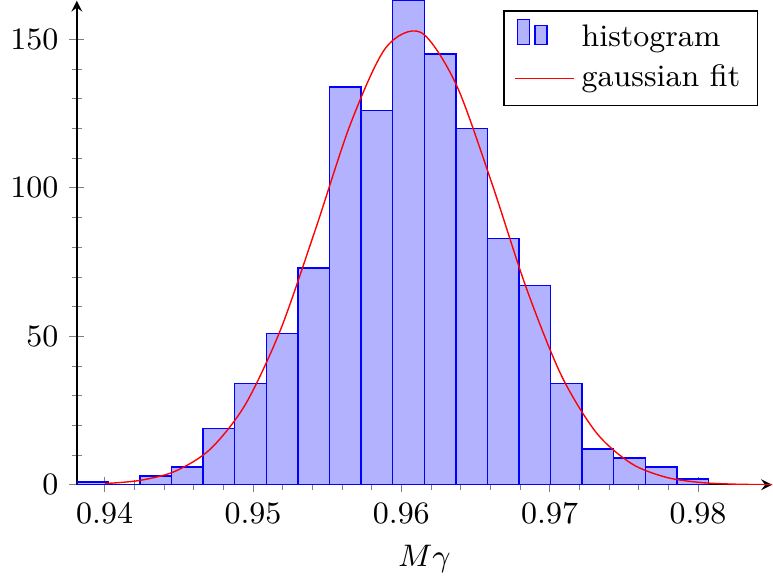}
	\caption{Histogram for $M=180$ and $x=0.255$. The ensemble consists of 1088 Lyapunov exponents.}
\label{fig2}
\end{figure}

{\it The fitting procedure.} Near the critical point in a system of finite width $M$, the LE $\Gamma$ is expected \cite{mackinnon1981scaling,mackinnon1983scaling,slevin2009} to exhibit the following scaling behavior:
\begin{align}
\label{ren_equ}
\Gamma = F_\Gamma[M^{1/\nu}u_0(x), f(M)\,u_1(x)],
\end{align}
where $F_\Gamma$ is a scaling function of the relevant field $u_0(x)$ and the leading irrelevant field $u_1(x)$. In the limit $M \to \infty$ the contribution of the irrelevant field should vanish, so $f(M)$ should decrease with $M$. If the field $u_1$ is truly irrelevant, we have $f(M)=M^y$ with a negative exponent $y<0$. Recently, it was suggested that one might need to include two irrelevant fields \cite{Nuding-Localization-2015}, or that the field $u_1$ can be marginally-irrelevant. The latter case would correspond to $f(M) = (\ln M)^{b}$ with some negative $b <0$ \cite{Amado-Numerical-2011, Nuding-Localization-2015}. In this work we assume only one irrelevant field $u_1$ characterized by $y<0$.

On the left hand side of \eqref{ren_equ} we use the numerical values of $\gamma$ extracted from $T_L$ for various combinations $x$ and $M$. The scaling function $\Gamma$ is expanded in its  arguments, and we assume that the scaling fields $u_i$ are polynomials in $x$. Since we do not have symmetry under $x \leftrightarrow -x$, and the critical point $x_c \neq 0$, we do not restrict polynomials $u_0(x)$, $u_1(x)$ to be even or odd. Then we get
\begin{align}
\label{expansion_in_fields}
	& F_\Gamma[u_0 M^{1/\nu},u_1 M^y] = \Gamma_{00} + \Gamma_{01} u_1 M^y + \Gamma_{20}u_0^2 M^{2/\nu}
\nonumber\\
	&  + \Gamma_{02}u_1^2M^{2y} + \Gamma_{21}u_0^2u_1M^{2/\nu}M^y + \Gamma_{03}u_1^3 M^{3y}+\dots,
\\
\label{fields_expanded}
 & u_0(x)=x + \sum_{k=2}^{m} a_{k}x^{k}, \quad \quad
 u_1(x)=1+\sum_{k=1}^{n} b_{k}x^{k} .
\end{align}
Because of ambiguity in the overall scaling of the fields, the
leading coefficient in \eqref{fields_expanded} can be chosen to be 1.

The critical exponents $\nu$ and $y$, and the critical amplitude ratio
$\Gamma_c \equiv \Gamma_{00}$ are the most interesting universal characteristics of the IQH transition. The latter is related to one of the multifractal exponents
by $\Gamma_c = \pi (\alpha_0 - 2)$ \cite{Janssen-Multifractal-1994, Dohmen-Disordered-1996}. These quantities, together with a finite number of expansion coefficients in Eqs.~(\ref{expansion_in_fields}) and (\ref{fields_expanded}) form sets $\Lambda = \{\nu, y, \Gamma_{ij}, a_k, b_l\}$ of the fitting parameters. The fits should use as few fitting parameters as possible while reproducing the data as well as possible, and we use several criteria to assess the quality of our fits. Details of our best fitting procedures are presented in the appendix~\ref{ensembleStats}.

{\it Results.} In Fig.~\ref{fig2} we present an example of a his\-to\-gram for the distribution of $\Gamma$ for $M=180$ and $x=0.255$. The distribution is fitted to a Gaussian, and the Gaussian fit is very accurate in full accord with Tutubalin's central limit theorem \cite{tutubalin1965limit}. As discussed in the appendix, each distribution for a given $(x,M)_a$ defines one data point and its error bars, as well as weights for the fitting procedures.

\begin{figure}[t]
	\centering
	\includegraphics{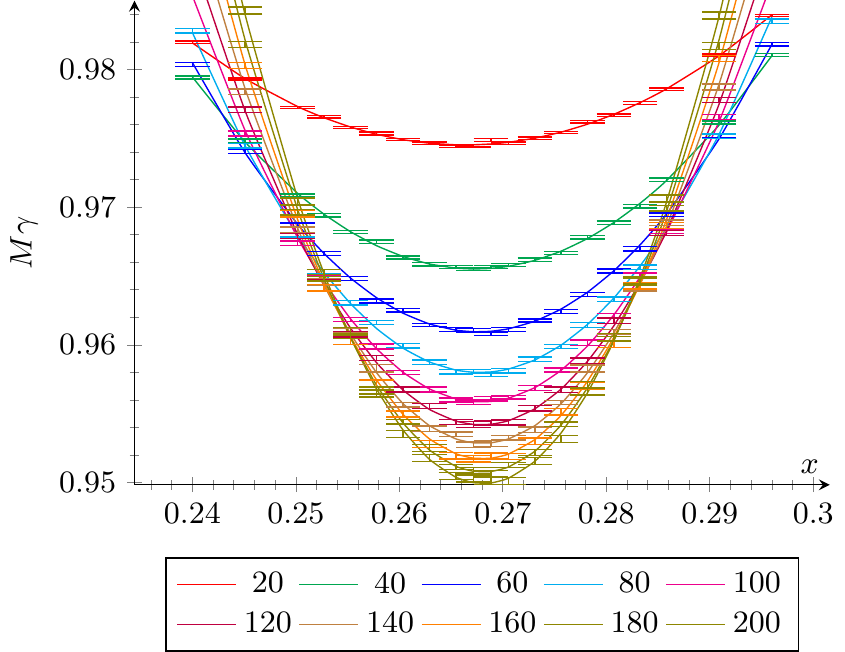}
	\caption{Results of the best fit for the Lyapunov exponents. The system widths $M$ are color-coded as indicated below the figure.}
\label{fig3}
\end{figure}

In Fig.~\ref{fig3} we plot the numerical data points for LEs, together with the scaling function $F_\Gamma$ that results from one of our two best fits. The two fits give the following values of the critical parameters (and the 95\% confidence bounds):
\begin{align}
\nu &= 2.658 & (2.612, 2.704), \\
y &= -0.1511 & (-0.4307, 0.1284), \\
\Gamma_c &= 0.9166 & (0.884, 0.9493),
\end{align}
and
\begin{align}
\nu &= 2.659 & (2.614, 2.704), \\
y &= -0.07007 & (-0.1625, 0.02232), \\
\Gamma_c&= 1.02 & (0.5593, 1.481).
\end{align}
Other fitting parameters $\Lambda$ are presented in the appendix.

{\it Conclusions and outlook.} We have studied the integer quantum Hall transition in the network model on the Kagome lattice (NMKL). We have argued that the model should exhibit critical properties that are the same as the Chalker-Coddington (CC) model on a square lattice. We simulated the NKML numerically using the transfer matrix approach, and obtained a number of critical properties, including the value $\nu= 2.658 \pm 0.046$ for the localization length exponent. This result is close to the standard CC model value obtained in a number of recent papers \cite{Slevin-Critical-2009, Obuse-Conformal-2010, Amado-Numerical-2011, Obuse-Finite-2012, Slevin-Finite-2012, Nuding-Localization-2015}. This indicates that the universality class of the transition in NMKL is the same as in the CC model, in spite of the presence of non-zero, periodically distributed curvature. Such regular, non-fluctuating curvature background turns out to be irrelevant and does not change the critical behavior. In contrast, in our previous papers \cite{Gruzberg-Geometrically-2017, Kluemper-Nuding-Sedrakyan-2019} we considered models with random, fluctuating curvatures, and found that their critical properties were distinct from those of the CC model, implying relevance of geometric disorder.

{\it Acknowledgments.}  The authors gratefully acknowledge the funding of this
project by computing time provided by the Paderborn Center for Parallel
Computing (PC2).  The work of A.~S.\ was partially supported by ARC grants
18T-1C153 and 18RF-039.  A.~K.\ is grateful to DFG (Deutsche
Forschungsgemeinschaft) for financial support in the framework of the research
unit FOR 2316.

\appendix

\section{Solution of the periodic network model on Kagome lattice}
\label{clean}

The network model on the Kagome lattice (NMKL) has a $\mathbb{Z}_3$ spectral symmetry, analogous to the $\mathbb{Z}_4$ spectral symmetry of the Chalker-Coddington (CC) model on a square lattice~\footnote{F. Evers, unpublished (2017); M. R. Zirnbauer, Nucl.Phys. B {\bf 941}, 458 (2019).}. Indeed, the Hilbert space of the network decomposes as ${\cal H} = \bigoplus_{k=0}^2 {\cal H}_k$. The subspace ${\cal H}_k$ is spanned by the states $|l\rangle$ on all links $l$ propagating at angles $k\pi/3$ and $k\pi/3 + \pi$ relative to the horizontal direction. It is clear that the operator $\cal U$ maps ${\cal H}_k$ to ${\cal H}_{k+1}$, where $k$ is taken modulo 3. Thus, if we order basis vectors appropriately, $\cal U$ becomes a block matrix
\begin{align}
{\cal U} &= \begin{pmatrix}
              0 & 0 & {\cal U}_0 \\
              {\cal U}_1 & 0 & 0 \\
              0 & {\cal U}_2 & 0
            \end{pmatrix},
\label{U-blocks}
\end{align}
where ${\cal U}_k: {\cal H}_{k-1} \to {\cal H}_k$. Using known formulas for determinants of block matrices, we can write the characteristic polynomial of $\cal U$ as
\begin{align}
\det ({\cal U} - \lambda) &= \det({\cal U}_2 {\cal U}_1 {\cal U}_0 -
\lambda^3).
\end{align}
Then the spectrum of $\cal U$ is obtained by taking cube roots of the eigenvalues of  ${\cal U}_2 {\cal U}_1 {\cal U}_0$.

\begin{figure}
\centering
\includegraphics[width=\columnwidth]{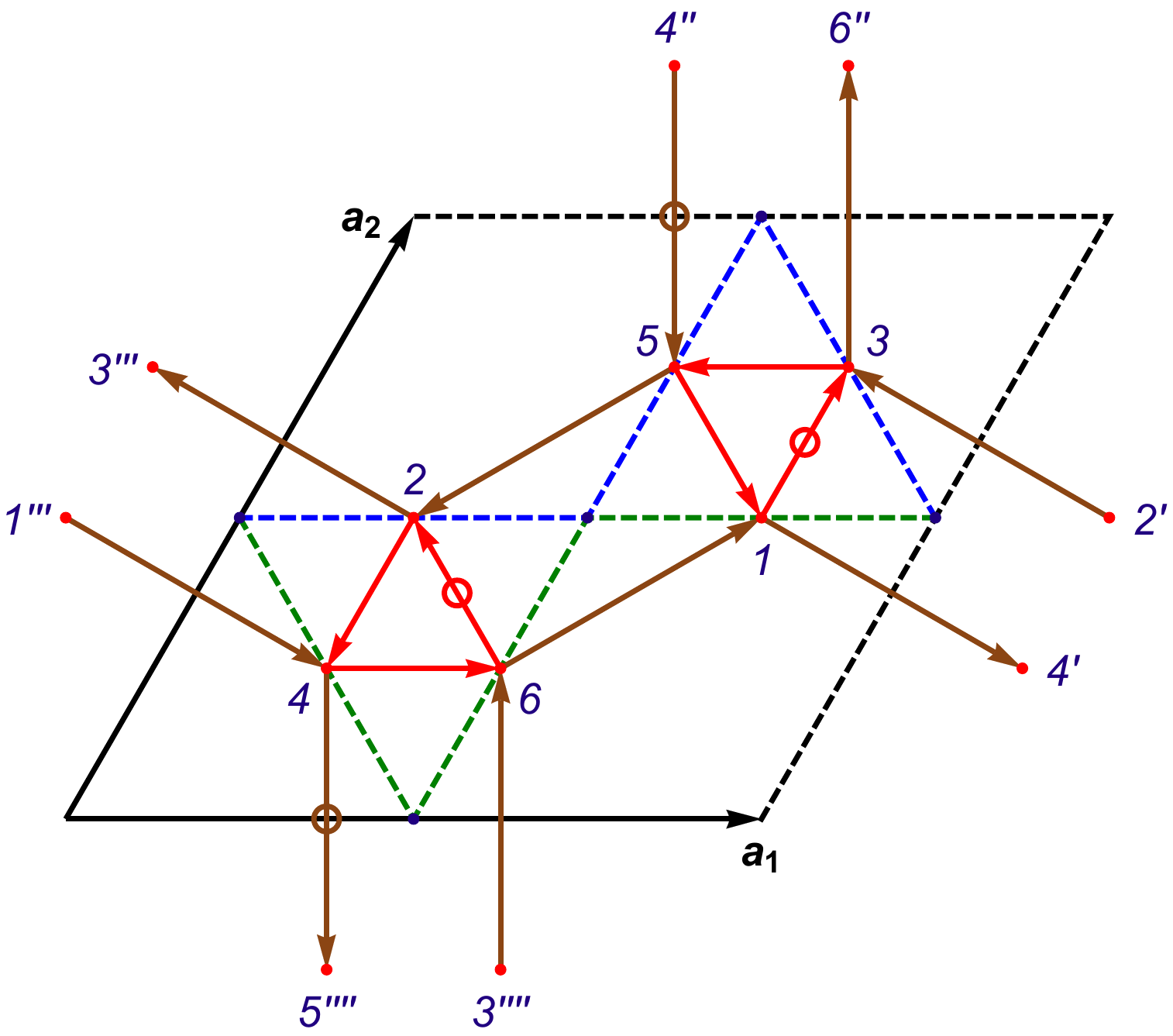}
\caption{A unit cell of the Kagome lattice and its edge graph. Red edges have amplitude $r = \cos \theta$, brown edges have amplitude $t = \sin \theta$ (with a factor of $-1$ acquired on the edges marked with a circle). In addition, an electron acquires a phase associated with the link of the NMKL onto which it travels.}
\label{gfx:kagomeunitcell}
\end{figure}

Let us now consider the NMKL without randomness. In this case the network is periodic with two lattice vectors whose $(x,y)$ components are (see Fig.~\ref{gfx:kagomeunitcell})
\begin{align}
\mathbf{a}_1 &= a (1,0), & \mathbf{a}_2 &= a (1/2,\sqrt{3}/2).
\end{align}
We parametrize the scattering amplitudes as
\begin{align}
r &= \cos \theta, & t &= \sin \theta.
\end{align}

The operator $\cal U$ can be viewed as describing a hopping of a particle on the directed edge graph of the Kagome lattice (also called median lattice). The edge graph has one vertex in the middle of every link of the Kagome network, and the vertices are connected by a directed edge when they correspond to two links of the network that meet at the same scattering node. In the hopping representation the hopping amplitudes for the longer edges of the edge graph (which lie inside the hexagons of the Kagome lattice and correspond to right turns on the original network) are $\pm t$, while the amplitudes for the shorter edges are $\pm r$. In addition, the particle acquires a phase associated to the link onto which it hops.

The operator $\cal U$ can be diagonalized going to the momentum space, where it becomes a $6 \times 6$ matrix ${\cal U}(\mathbf{k})$ with the block structure as in Eq.~(\ref{U-blocks}), where the blocks ${\cal U}_k(\mathbf{k})$ are $2 \times 2$ matrices. If we label the vertices of the edge graph within a unit cell as in Fig.~\ref{gfx:kagomeunitcell}, the blocks ${\cal U}_k(\mathbf{k})$ have the following explicit form:
\begin{align}
{\cal U}_0(\mathbf{k})
&= \begin{pmatrix} e^{i \phi_1} & 0 \\ 0 & e^{i \phi_2}
   \end{pmatrix}
   \begin{pmatrix} r & t \\ t & -r
   \end{pmatrix},
   \nonumber \\
{\cal U}_1(\mathbf{k})
&= \begin{pmatrix} e^{i \phi_3} & 0 \\ 0 & e^{i \phi_4}
   \end{pmatrix}
   \begin{pmatrix} -r & e^{i \mathbf{k} \cdot \mathbf{a}_1} t \\
   e^{-i \mathbf{k} \cdot \mathbf{a}_1} t & r
   \end{pmatrix},
   \nonumber \\
{\cal U}_2(\mathbf{k})
&= \begin{pmatrix} e^{i \phi_5} & 0 \\ 0 & e^{i \phi_6}
   \end{pmatrix}
   \begin{pmatrix} r & -e^{i \mathbf{k} \cdot \mathbf{a}_2} t \\
   e^{-i \mathbf{k} \cdot \mathbf{a}_2} t & r
   \end{pmatrix}.
\end{align}
We find
\begin{align}
\det {\cal U}(\mathbf{k}) &= \prod_{k=0}^{2} \det {\cal U}_k(\mathbf{k}) = e^{i \Phi},
&
\Phi &= \sum_{l=1}^{6} \phi_l.
\end{align}

Then ${\cal U}'(\mathbf{k}) \equiv e^{-i \Phi/6} {\cal U}(\mathbf{k})$ is a special unitary matrix with blocks ${\cal U}'_k(\mathbf{k})$. Let us now denote
\begin{align}
W \equiv {\cal U}'_2(\mathbf{k}) {\cal U}'_1(\mathbf{k}) {\cal U}'_0(\mathbf{k})
= \begin{pmatrix}
    a & b \\
    -b^* & a^*
  \end{pmatrix}.
\end{align}
This is a special unitary matrix whose eigenvalues can be written $e^{i \chi}$ and $e^{-i \chi}$ for $0 \leq \chi \leq \pi$. Due to the invariance of the trace under similarity transformations we have
\begin{align}
2 \cos \chi = \text{tr}\, W = a + a^* = 2 \, \text{Re} \, a.
\label{cos-chi-1}
\end{align}
The explicit computation gives
\begin{align}
a &= - e^{i \Delta \phi} \cos^3 \theta
+ (e^{i q_1} + e^{i q_2} + e^{i(q_2 - q_1 + \Delta \phi)}) \sin^2 \theta \cos \theta,
\label{a}
\end{align}
where we have introduced the notations
\begin{align}
2 \Delta \phi &= \phi_1 + \phi_3 + \phi_5 - (\phi_2 + \phi_4 + \phi_6),
\nonumber \\
q_1 &= \mathbf{k} \cdot \mathbf{a}_1 + \Delta \phi - \phi_1 + \phi_2,
\nonumber \\
q_2 &= \mathbf{k} \cdot \mathbf{a}_2 - \Delta \phi + \phi_5 - \phi_6 + \pi.
\end{align}

Taking the real part of Eq.~(\ref{a}) gives
\begin{align}
\cos \chi &=
- A(q_1, q_2, \Delta \phi) \cos^3 \theta + B(q_1, q_2, \Delta \phi) \cos \theta,
\label{cos-chi-main}
\end{align}
where
\begin{align}
B(q_1, q_2, \Delta \phi) &= \cos q_1 + \cos q_2 + \cos(q_2 - q_1 + \Delta \phi),
\nonumber \\
A(q_1, q_2, \Delta \phi) &= B(q_1, q_2, \Delta \phi) + \cos (\Delta \phi)
\nonumber\\
&= 4 \cos \frac{q_1 - q_2}{2} \cos \frac{q_1 - \Delta \phi}{2}
\cos \frac{q_2 + \Delta \phi}{2}.
\end{align}

Equation (\ref{cos-chi-main}) determines $\chi(\mathbf{k}, \theta, \{\phi\})$ in the Brillouin zone  as a function of the scattering angle $\theta$ and the phases $\phi_l$. The six quasi-energies $\varepsilon_{s,m}$ of the original evolution matrix ${\cal U}(\mathbf{k})$ are
\begin{align}
\epsilon_{s,m} &= s \frac{\chi}{3} + m \frac{2\pi}{3} + \frac{\Phi}{6},
& s &= \pm1, & m &= 0,1,2.
\label{bands}
\end{align}
For generic values of parameters, the six bands $\varepsilon_{s,m}$ are non-degenerate, and the system is gapped. However, for any values of the phases $\phi_l$ there is a critical value $\theta_c$ for which the six bands touch pairwise at a single degeneracy point $\mathbf{k}_0$ in the Brillouin zone (BZ). In the vicinity of $\mathbf{k}_0$ the two touching bands have the shape of a Dirac cone. Let us see how this all comes about.

First of all for the two bands $\varepsilon_{s_1,m_1}$ and $\varepsilon_{s_2,m_2}$ to touch, the signs $s_1$ and $s_2$ must be opposite to represent the bands dispersing away from the degeneracy point in opposite directions. We equate $\varepsilon_{1,m_1} = \varepsilon_{-1,m_2}$ and find $\chi = \pi m$, where $m = m_1 - m_2 \mod 2$.

Equation (\ref{cos-chi-main}) is unchanged when the signs of $\cos \chi$ and of $\cos \theta$ are flipped simultaneously. The sign of $\cos \theta$ can be flipped by a gauge transformation, so it is sufficient to choose the value $\chi = 0$ (which turns out to correspond to $\cos \theta_c = 1/2$ for $\Delta\phi = 0$). Then we get a cubic equation for $X \equiv \cos \theta$
\begin{align}
A X^3 - B X + 1 = 0.
\label{cubic-equation}
\end{align}
All three roots $X_k$ ($k = 0,1,2$) of this equation can be explicitly written in the trigonometric form as
\begin{align}
X_k &= 2 \sqrt{\frac{B}{3A}} \cos \Big(\frac{1}{3} \arccos \Big(- \frac{3}{2B} \sqrt{\frac{3A}{B}} \Big) - \frac{2\pi k}{3} \Big).
\end{align}

The nature of these roots depends on the value of the discriminant $D = 4A B^3 - 27 A^2$ of the cubic equation. One of the roots is always real and does not correspond to criticality (or lies outside the physical range of values $[-1,1]$ for $\cos \theta_c$). The other two roots are real when $D < 0$ and complex when $D > 0$. The critical point corresponds to the values of parameters when a real double root appears in addition to the un-physical real root. This happens when the discriminant vanishes:
\begin{align}
4 A B^3 - 27 A^2 = 0.
\label{D-zero}
\end{align}

Consider now the symmetries of Eq.~(\ref{cubic-equation}). The values of $A(q_1, q_2, \Delta \phi)$ and $B(q_1, q_2, \Delta \phi)$ are invariant under the exchange $q_1 \leftrightarrow -q_2$. Geometrically, this is the reflection across the line $q_1 + q_2 = 0$ in the BZ. We expect that the criticality (band touching) should happen only at one point in the BZ. Then this point must be on the reflection line, so that $q_2 = - q_1 = -q$. On this line
\begin{align}
A(q,\Delta \phi) &= 2 \cos q + \cos(2q - \Delta \phi) + \cos \Delta \phi
\nonumber \\
&= 4 \cos q \cos^2 \frac{q - \Delta \phi}{2}
\nonumber \\
B(q,\Delta \phi) &= 2 \cos q + \cos(2q - \Delta \phi).
\end{align}
These expressions are periodic in $\Delta \phi$ with period $2\pi$, and invariant under the simultaneous sign flip of $q$ and $\Delta \phi$. Thus, we can analyze everything for $\Delta\phi \in [0,\pi]$.

The trivial solution of Eq.~(\ref{D-zero}), $A = 0$ ($B = - \cos \Delta \phi$), corresponds to an un-physical value $\cos \theta_c = 1/\cos \Delta \phi$. The relevant non-trivial solution is achieved at $q = \Delta\phi/3$ when $A = 4 \cos^3 (\Delta\phi/3)$, $B = 3 \cos (\Delta\phi/3)$. The corresponding critical turning amplitude is given by
\begin{align}
r_c &= \cos \theta_c = \frac{1}{2} \sec \frac{\Delta\phi}{3},
\end{align}
and the degeneracy point $\mathbf{k}_0$ satisfies
\begin{align}
\mathbf{k}_0 \cdot \mathbf{a}_1 &= - \frac{2}{3}\Delta \phi +  \phi_1 - \phi_2,
\nonumber\\
\mathbf{k}_0 \cdot \mathbf{a}_2 &= \frac{2}{3}\Delta \phi - \phi_5 + \phi_6 - \pi.
\end{align}

\begin{figure}
\centering
\includegraphics[width=0.8\columnwidth]{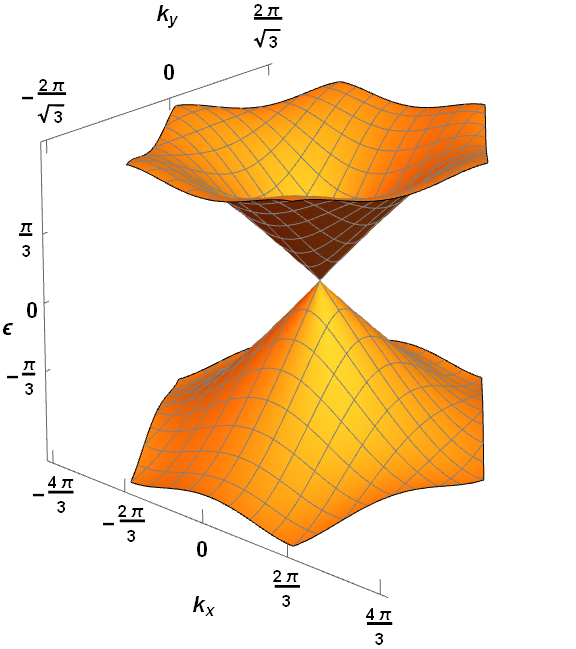}
\caption{The two bands $\epsilon_{\pm,0}$ (see Eq.~(\ref{bands}) in the text) touching at $\epsilon = 0$ and forming a Dirac cone at $\mathbf{k} = 0$.}
\label{gfx:dirac-cone}
\end{figure}

Expanding Eq.~(\ref{cos-chi-main}) near the degeneracy point $\chi = \pi + \delta \chi$, $\mathbf{k} = \mathbf{k}_0 + \delta \mathbf{k}$, we get in leading order the Dirac cone
\begin{align}
\delta \chi^2 &= \frac{3}{4}\sin^2 \theta_c (\delta k_x^2 + \delta k_y^2).
\end{align}
In particular, when all $\phi_l = 0$, the Dirac point $\mathbf{k}_0 = 0$ is in the center of the BZ (see Fig.~\ref{gfx:dirac-cone}) and the critical amplitude
\begin{align}
r_c &= \cos \theta_c = \frac{1}{2}.
\end{align}

\section{Details of fitting procedure}
\label{ensembleStats}

\begin{table*}[t]
  \centering
	\begin{tabular}{|r| *{20}{c} |}
		\hline
		\bm{$M$} & \multicolumn{20}{c}{\bm{$x$}}  \vline \\
		&  0.24 &  0.245 &  0.250 &  0.253 &  0.255 &  0.258 &  0.260 &  0.263 &  0.265 &  0.267 &  0.269 &  0.271 &  0.273 &  0.276 &  0.278 &  0.281 &  0.283 &  0.286 &  0.291 &  0.296 \\
		\hline
		20 & 624 & 624 & 624 & 416 & 624 & 368 & 624 & 416 & 624 & 416 & 432 & 624 & 416 & 816 & 416 & 624 & 416 & 624 & 608 & 624  \\
		40 & 592 & 592 & 608 & 576 & 560 & 544 & 576 & 560 & 608 & 576 & 576 & 592 & 560 & 544 & 496 & 592 & 528 & 560 & 608 & 592  \\
		60 & 624 & 624 & 624 & 624 & 624 & 608 & 624 & 832 & 624 & 832 & 832 & 624 & 832 & 624 & 592 & 624 & 624 & 624 & 624 & 624  \\
		80 & 640 & 640 & 640 & 624 & 640 & 576 & 624 & 624 & 624 & 624 & 624 & 624 & 624 & 608 & 576 & 624 & 624 & 624 & 624 & 624  \\
		100 & 624 & 624 & 816 & 624 & 1024 & 608 & 1040 & 624 & 1024 & 624 & 624 & 1040 & 624 & 1008 & 592 & 1040 & 624 & 816 & 624 & 608  \\
		120 & 624 & 624 & 624 & 624 & 624 & 608 & 624 & 608 & 624 & 624 & 624 & 624 & 608 & 624 & 576 & 624 & 624 & 624 & 624 & 624  \\
		140 & 608 & 624 & 816 & 528 & 1008 & 384 & 1024 & 624 & 1024 & 624 & 624 & 1040 & 624 & 1040 & 416 & 1040 & 528 & 832 & 624 & 624  \\
		160 & 624 & 624 & 624 & 640 & 624 & 544 & 624 & 608 & 624 & 576 & 624 & 608 & 592 & 624 & 592 & 624 & 608 & 624 & 624 & 624  \\
		180 & 640 & 688 & 880 & 640 & 1088 & 560 & 1088 & 624 & 1056 & 688 & 656 & 1040 & 704 & 1088 & 624 & 1072 & 624 & 848 & 640 & 672  \\
		200 & 624 & 608 & 608 & 624 & 624 & 576 & 624 & 624 & 624 & 624 & 624 & 624 & 624 & 624 & 592 & 624 & 608 & 624 & 624 & 624  \\
		\hline
	\end{tabular} \\
  \caption{Numbers $N_\alpha$ of Lyapunov exponents in ensembles created for each pair $(x,M)$.}
  \label{Table:N-ensembles}
\end{table*}
Table \ref{Table:N-ensembles} shows the numbers $N_\alpha$ of the smallest LEs generated for each of the 200 pairs $(x,M)_a$ that we studied. All ensembles of the Lyapunov exponents were obtained in systems of length $L=5\times 10^6$.

{\it Errors and weights.} For each ensemble $a$ of the dimensionless LEs $\Gamma$ corresponding to a given combination $(x,M)_a$ we determine its mean and variance
\begin{align}
\bar{\Gamma}_a &= \frac{1}{N_a} \sum_{i=1}^{N_\alpha} \Gamma_i,
&
\sigma_a^2 &= \frac{1}{N_a-1} \sum_{i=1}^{N_a} (\Gamma_i - \bar{\Gamma}_a)^2.
\end{align}
For example, for the distribution shown in Fig.~\ref{fig2} these turn out to be
\begin{align}
\bar{\Gamma} &= 0.96066, & \sigma &= 0.0060828.
\label{mean+SD}
\end{align}
To check how close our distribution is to the Gaussian predicted by Tutubalin's theorem, we fitted it to $p(\Gamma) = a e^{-(\Gamma - \mu)^2/2s^2}$, and obtained the following parameters (with 95\% confidence bounds)
\begin{align}
\mu &= 0.9607, & (0.9603, 0.9611),
\nonumber \\
s &= 0.006, & (0.005607, 0.006394),
\nonumber \\
a &= 2.305, & (2.174, 2.435).
\end{align}
The resulting probability density is shown as a red curve in Fig.~\ref{fig2}. The closeness of the pairs $\bar{Gamma}$ and $\mu$ and $\sigma$ and $s$ demonstrates the quality of the Gaussian fit.

If we now view each $\Gamma_i$ in a given ensemble $a$ as drawn for the same distribution with variance $\sigma_a^2$, then the mean $\bar{\Gamma}_a$ is also a random variable whose variance is
\begin{align}
\bar{\sigma}_a^2 = \sigma_a^2/N_a.
\end{align}
The corresponding standard deviation $\bar{\sigma}_a = \sigma_a/N_a^{1/2}$ is used as the error for the data points $\bar{\Gamma}_a$ for each of the 200 ensembles shown in Fig.~\ref{fig3}. The inverse variances $\sigma_a^{-2}$ and $\bar{\sigma}_a^{-2}$ are also used as weights for the subsequent least square fits of the numerical data to the scaling function in Eq.~(\ref{ren_equ}).

Next, we perform a weighted nonlinear least square fit based on a trust region algorithm with specified regions for each fitting parameter. The resulting parameters are used in a subsequent weighted nonlinear least square fit based on a Levenberg-Marquardt algorithm. Here no limits are imposed on the fit parameters. The last step is repeated until the resulting fitting parameters stop changing.

We used two schemes to do the least square minimization. We define two quantities:
\begin{align}
\chi^2_1[\Lambda] &= \sum_{a=1}^{N_\text{ens}} \frac{(\bar{\Gamma}_a -
F_\Gamma[\Lambda,(x,M)_a])^2}{\bar{\sigma}_a^2},
\label{chi-squared-1}
\\
\chi^2_2[\Lambda] &= \sum_{a=1}^{N_\text{ens}} \sum_{i=1}^{N_a} \frac{(\Gamma_i -
F_\Gamma[\Lambda,(x,M)_a])^2}{\sigma_a^2}.
\label{chi-squared-2}
\end{align}
The first quantity, $\chi^2_1$, contains $N_\text{ens} = 200$ in the sum. It involves the mean $\bar{\Gamma}_a$ from each ensemble, and uses the inverse of its variance $\bar{\sigma}_a^{-2}$ as the weight. On the other hand, the second quantity $\chi^2_2$, contains all $N_\text{LE} = 130896$ individual LEs $\Gamma_i$ and their variances $\sigma_a^2$ that depend only on the ensemble $a$ to which each LE $\Gamma_i$ belongs. In the two schemes, both quantities $\chi^2_1$ and $\chi^2_2$ are minimized to obtain the values of the optimal parameters $\Lambda$.

{\it Evaluation of fits}. Next we evaluate the quality of the fits. We present several methods to do this. The most important one is the $\chi^2$ test, where $\chi^2$ is the actual minimum of $\chi^2[\Lambda]$ achieved in Eq.~(\ref{chi-squared-1}) or (\ref{chi-squared-2}).

As our fits contain many data points with the same pairs $(x,M)$, $\chi^2=0$ is not possible. In fact, the individual terms in the sums in Eqs.~(\ref{chi-squared-1}) and (\ref{chi-squared-2}) are designed to be of order 1. Therefore, we expect $\chi^2_1 \sim N_\text{ens} = 200$ and $\chi^2_2 \sim N_\text{LE} = 130896$. One usually considers the ratio $\chi^2$/\emph{dof}, where \emph{dof}, the number of degrees of freedom, is the difference between the number of terms in the sums in Eqs.~(\ref{chi-squared-1}) and (\ref{chi-squared-2}) and the number of the fitting parameters in the set $\Lambda$. The expected value for the ratio $\chi^2$/\emph{dof} is 1 for an ideal fit.

Deviations from 1 are evaluated with the cumulative probability $P(\tilde\chi^2 < \chi^2)$ which is the probability of observing -- just for statistical reasons -- a sample statistic with a smaller $\chi^2$ value than in our fit. A small value of $P$, i.e.~a large value of the complement $Q:=1-P$ is taken as an indication of a good fit. However, values of $P$ lower than $1/2$ indicate problems in the estimation of the error bars of the individual data points.

Another criterion is based on the width of \emph{confidence intervals}. This quantifies the quality of the prediction for a single parameter. We use 95\% confidence intervals, which means that for repeated independent generation of the same amount of data and application of the same kind of data analysis, the resulting confidence intervals contain the true parameter values in 95\% of the cases.

The last criterion we present is the sum of \emph{residuals}. In the two fitting schemes this is given by
\begin{align}
\text{res} &= \sum_{a=1}^{N_\text{ens}} \text{res}_a, &
\text{res}_a &= \bar{\Gamma}_a - F_\Gamma[\Lambda,(x,M)_a],
\\
\text{res} &= \sum_{a=1}^{N_\text{ens}} \sum_{i=1}^{N_a} \text{res}_i, &
\text{res}_i &= \Gamma_i - F_\Gamma[\Lambda,(x,M)_a].
\end{align}
In a good fit the sum of residuals should be small compared to \emph{dof}, while the individual residuals $\text{res}_a$ and $\text{res}_i$ plotted as functions of $a$ and $i$ should fluctuate in sign and look like noise around zero. If the residuals do not scatter around zero, it indicates that the fit function is not correct.

Now we present the results of our best two fits. They have been obtained by expanding $\Gamma$ up to second order in $u_0$, and first order in $u_1$ \eqref{expansion_in_fields}, and expanding $u_0$ and $u_1$ up to the third order in $x$ \eqref{fields_expanded}. Thus, in both schemes the optimal number of parameters turned out to be 13.

For the fitting scheme using Eq.~(\ref{chi-squared-1}) we found the following fitting and goodness of fit parameters:
\vskip 3mm
Fitting parameters (confidence bounds 95\%):
\begin{align*}
	\hline \\[-2.5ex]
	\Gamma_{00} =\; &	\quad  0.9166 & (0.884, 0.9493)\\
	\Gamma_{01} =\; &	\quad 0.03359 & (-0.3693, 0.4365)\\
	\Gamma_{10} =\; &	-0.02351 & (-0.05937, 0.01234) \\
	\Gamma_{02} =\; &	\quad  0.0904 & (-0.2669, 0.4477)\\
	\Gamma_{11} =\; &	\quad  0.04551 & (0.0126, 0.07843) \\
	\Gamma_{20} =\; &	\quad  1.206 & (1.122, 1.29)\\
	a_2         =\; &	\quad  0.1916 & (0.05975, 0.3235)\\
	a_3         =\; &	\quad  3.064 & (-2.913, 9.04)\\
	b_1         =\; &	\quad  0.2062 & (-0.08682, 0.4992)\\
	b_2         =\; &	-9.163 & (-25.3, 6.974)\\
	b_3         =\; &	-41.8 & (-175.7, 92.07)\\
	\nu         =\; &	\quad 2.658 & (2.612, 2.704)\\
	y           =\; &	-0.1511 & (-0.4307, 0.1284)\\
	\hline
\end{align*}
\vskip -3mm
Goodness of fit parameters:
\begin{align*}
  \hline \\[-2.5ex]
	&\chi^2 && 204.1911 \\
	&\text{degrees of freedom (\emph{dof})} && 187 \\
	& \chi^2/\text{\emph{dof}} && 1.0919 \\
	& P && 0.81522 \\
	&\text{sum of residuals} && 0.14624 \\
  \hline
\end{align*}

In the second fitting scheme we used Eq.~(\ref{chi-squared-2}) and obtained the following results:
\vskip 3mm
Fitting parameters (confidence bounds 95\%):
\begin{align*}
	\hline \\[-2.5ex]
	\Gamma_{00} =\; &	\quad  1.02 & (0.5593, 1.481)\\
	\Gamma_{01} =\; &	-0.361 & (-1.722, 1)\\
	\Gamma_{10} =\; &	-0.04203 & (-0.09575, 0.01169) \\
	\Gamma_{02} =\; &	\quad  0.3761 & (-0.517, 1.269)\\
	\Gamma_{11} =\; &	\quad  0.056 & (-0.001508, 0.1135) \\
	\Gamma_{20} =\; &	\quad  1.208 & (1.125, 1.29)\\
	a_2         =\; &	\quad  0.1727 & (0.04725, 0.2982)\\
	a_3         =\; &	\quad  3.048 & (-2.634, 8.729)\\
	b_1         =\; &	\quad  0.124 & (-0.03855, 0.2866)\\
	b_2         =\; &	-4.273 & (-9.722, 1.176)\\
	b_3         =\; &	-9.875 & (-61.3, 41.55)\\
	\nu         =\; &	\quad 2.659 & (2.614, 2.704)\\
	y           =\; &	-0.07007 & (-0.1625, 0.02232)\\
	\hline \\
\end{align*}
\vskip -3mm
Goodness of fit parameters:
\begin{align*}
  \hline \\[-2.5ex]
	&\chi^2 && \quad\;  1.3094e+05 \\
	&\text{degrees of freedom (\emph{dof})} && \quad\;  130883 \\
	& \chi^2/\text{\emph{dof}} && \quad\;  1.0004 \\
	& P && \quad\;  0.54482 \\
	&\text{sum of residuals} && -32.1663 \\
  \hline
\end{align*}
Here $\chi^2/$\emph{dof} is close to 1 and the cumulative probability $P=0.52379$ is
close to $1/2$ implying a good fit result. The sum of residuals is small compared
to the degrees of freedom. In a plot the residuals are distributed around zero
by eye's measure. All this indicates that the fit is reliable and the data agree
with the scaling function well.


\nocite{Levine1984yg,Levine1984yf,Levine1984ye}

\bibliographystyle{my-refs}
\bibliography{Literaturverzeichnis,IQH,Anderson-localization-pre-AZ,
Anderson-localization-post-AZ,Scattering-networks,2D-quantum-gravity}

\end{document}